\documentclass[twocolumn,secnumarabic,amssymb, nobibnotes, aps, prd]{revtex4}
\usepackage{graphicx}

\begin{document}
\title{Quantum oscillations of the critical current of asymmetric superconducting rings and systems of the rings}
\author{V.L. Gurtovoi$^{1}$, M. Exarchos$^{2}$, R. Shaikhaidarov$^{2}$, V.N. Antonov$^{2}$, A.V. Nikulov$^{1}$, and V.A. Tulin$^{1}$}
\affiliation{$^{1}$Institute of Microelectronics Technology and High Purity Materials, Russian Academy of Sciences, 142432 Chernogolovka, Moscow District, RUSSIA. \\ $^{2}$ Physics Department, Royal Holloway University of London, Egham, Surrey TW20 0EX, UK} 
\begin{abstract} The quantum oscillations in magnetic field of the critical current of asymmetric superconducting rings with different widths of the half-rings are shifted to opposite sides for measurement in the opposite direction. The value of this shift found before equal half of the flux quantum for single ring with radius $2 \ \mu m$ has smaller value at lower critical current and for system of the rings with smaller radius. 
 \end{abstract}

\maketitle

\narrowtext

\section*{Introduction}

Quantum mechanics was developed for description of paradoxical phenomena observed on the atomic level with typical size $ < 1 \ nm = 10^{-9} \ m$. In our everyday word with typical size $ > 0.1 \ mm = 10^{-4} \ m$ quantum phenomena are not observed. One may say that nanostructures appertain to a region of boundary between the quantum and classical worlds. For example, according to the Bohr' quantization of the angular momentum of electron $rp = rmv = \hbar n$ postulated in 1913 the energy difference between permitted states $E_{n+1,n} = mv_{n+1}^{2}/2 - mv_{n}^{2}/2 \approx  \hbar^{2}/2mr^{2} \approx  5.5 10^{-21}  \ J \approx  k_{B} \ 400 \ K$ at the orbit radius $r \approx  1 \ nm$ and  $E_{n+1,n} \approx  k_{B} \ 0.0004 \ K$ at $r \approx  1 \ \mu m = 1000 \ nm$. Therefore atom orbits with $r < 1 nm$ are stable at the room temperature $T \approx  300 K$ whereas the persistent current, the quantum phenomena connected with the Bohr' quantization, can be observed in semiconductor [1] and normal metal [2] ring with $r \approx  1 \ \mu m$ only at very low temperature $T < 1 \ K$. In the rings with radius $r \approx 10 \ nm$  made recently [3] this phenomenon was observed up to $T = 4.2 \ K$.

\begin{figure}[]
\includegraphics{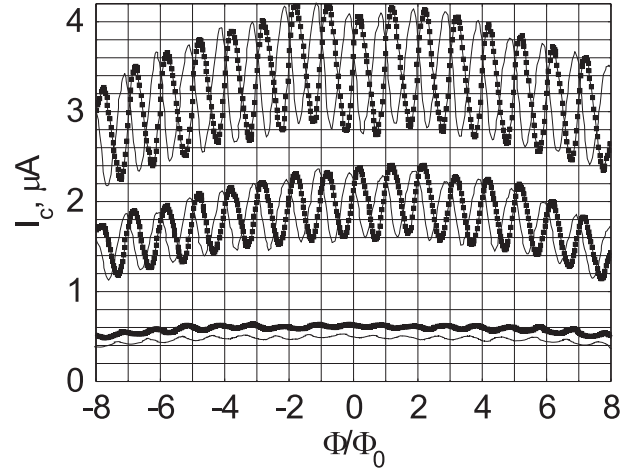}
\caption{\label{fig:epsart} Magnetic field dependencies of the critical current of a single asymmetric ring with radius $r \approx  2 \ \mu m$ and widths of the half-rings $w_{w} = 0.3 \ \mu m$ and $w_{n} = 0.2 \ \mu m$ measured in the positive $I_{c+}$ (solid lines) and negative $I_{c-}$ (black squares) directions at $T = 0.986T_{c}$ ($I_{c}(T) = 4.3 \ \mu A$); $T = 0.990T_{c}$ ($I_{c}(T) = 2.5 \ \mu A$) and $T = 0.996T_{c}$ ($I_{c}(T) = 0.6 \ \mu A$). $T_{c} = 1.294 \ K$. The $I_{c+}(\Phi/\Phi_{0})$ dependence for $T = 0.996T_{c}$ is vertically shifted on $ - 0.1 \ \mu A$. The period of oscillations in magnetic field $H_{0} = \Phi_{0}/S = 1.44 \ Oe$ corresponds to the rings area $S = \pi r^{2} = 14.4 \mu m^{2}$ and the radius $r \approx  2.1 \ \mu m$. }
\end{figure}

\begin{figure}
\includegraphics{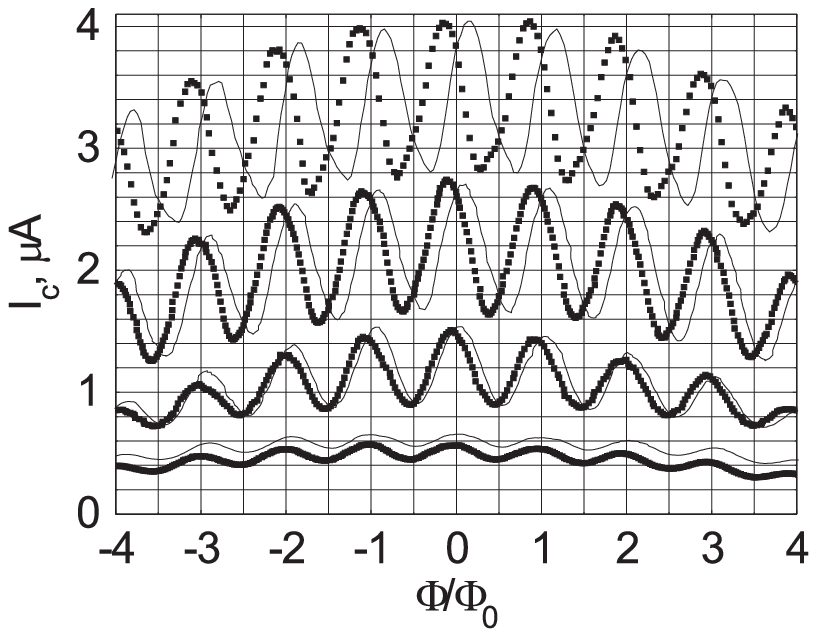}
\caption{\label{fig:epsart} Magnetic field dependencies of the critical current of an system of 110 asymmetric ring with radius $r \approx  1 \ \mu m$ and widths of the half-rings  $w_{w} = 0.4 \ \mu m$ and $w_{n} = 0.2 \ \mu m$ measured in the positive $I_{c+}$ (solid lines) and negative $I_{c-}$ (black squares) directions at $T = 0.968T_{c}$ ($I_{c}(T) = 4.1 \ \mu A$); $T = 0.975T_{c}$ ($I_{c}(T) = 2.8 \ \mu A$); $T = 0.981T_{c}$ ($I_{c}(T) = 1.6 \ \mu A$) and $T = 0.989T_{c}$ ($I_{c}(T) = 0.7 \ \mu A$). $T_{c} = 1.345 \ K$. The $I_{c-}(\Phi/\Phi_{0})$ dependence for $T = 0.996T_{c}$ is vertically shifted on $ - 0.1 \ \mu A$. The period of oscillations in magnetic field $H_{0} = \Phi_{0}/S \approx  5.2 \ Oe$ corresponds to the rings area $S \approx  4 \mu m^{2}$ and the radius $r \approx  1.1 \ \mu m$.  }
\end{figure}

\begin{figure}
\includegraphics{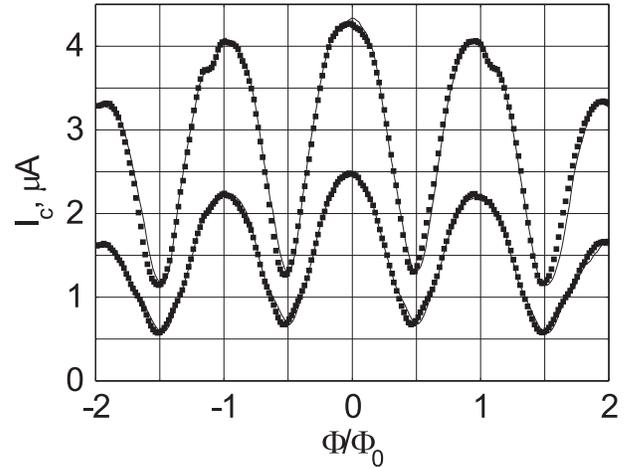}
\caption{\label{fig:epsart} Magnetic field dependencies of the critical current of an system of 667 asymmetric ring with radius $r \approx  0.5 \ \mu m$ and widths of the half-rings $w_{w} = 0.2 \ \mu m$ and $w_{n} = 0.15 \ \mu m$ measured in the positive $I_{c+}$  (solid lines) and negative $I_{c-}$  (black squares) directions at $T = 0.955T_{c}$ ($I_{c}(T) = 4.3 \ \mu A$) and $T = 0.960T_{c}$ ($I_{c}(T) = 2.5 \ \mu A$). $T_{c} = 1.327 \ K$. The period of oscillations in magnetic field $H_{0} = \Phi_{0}/S = 22.6 \ Oe$ corresponds to the rings area $S \approx  0.92 \mu m^{2}$ and the radius $r \approx  0.54 \ \mu m$. }
\end{figure}

The persistent current is observed because of the influence of the magnetic vector potential $A$ on the phase $\varphi $ of the wave function $\Psi = |\Psi |\exp{i\varphi }$ the gradient of which $\bigtriangledown \varphi $ is proportional to the canonical momentum $\hbar \bigtriangledown \varphi = p = mv + qA$ of a particle with the mass $m$ and the charge $q$. The minimum permitted value of the velocity $ \oint_{l}dl v = m^{-1} \oint_{l}dl (\hbar \bigtriangledown \varphi - qA) = m^{-1} (2\pi \hbar n - q\Phi) $ should be periodical function of the magnetic flux $\Phi $ inside the ring since the $\Psi = |\Psi |\exp{i\varphi }$ must be single-valued at any point of the ring circumference. The period of oscillations in magnetic field $H_{0}$ corresponds to the flux quantum $H_{0}S = \Phi_{0} = 2\pi \hbar /q$ inside the ring with the area $S = \pi r^{2}$.

\section {Quantum oscillations of the persistent current}
The energy difference between permitted states of any real superconductor ring $\Delta E_{n+1,n} >> k_{B}T$ since superconducting condensate moves as whole and superconducting pairs can not change the integer quantum number $n$ individually [4]. Therefore the persistent current $I_{p} = sj_{p} = sqn_{s}v$ is observed even above superconducting transition [5]. The $n - \Phi/\Phi_{0}$ value in the relation 
$$I_{p} = I_{p,A}2(n - \frac{\Phi}{\Phi_{0}}) \eqno{(1)}$$ 
for the permitted value of the persistent current corresponding to minimum energy $E_{n} \propto  (n - \Phi /\Phi _{0} )^{2}$ changes between -1/2 and 1/2 with the $\Phi $ variation from $(n + 0.5)\Phi_{0}$ to $(n + 1 + 0.5)\Phi_{0}$ and jumps from 1/2 to -1/2 at $(n + 0.5)\Phi_{0}$ [6]. The state with minimum energy gives predominant contribution to the thermodynamic average value $\overline{I_{p}}$. This value equals zero $\overline{I_{p}} = 0$ not only at $\Phi = n \Phi_{0}$ but also at $\Phi = (n + 0.5)\Phi_{0}$ where the persistent current $I_{p}$ have equal magnitude and opposite direction in the two states $n - \Phi/\Phi_{0} = \pm 1/2$. In agreement with (1) the magnetization $M \propto \overline{I_{p}}$ [5] and the rectified voltage $V_{dc} \propto \overline{I_{p}}$ [7,8] equal zero at $\Phi = n\Phi_{0}$ and $\Phi = (n + 0.5)\Phi_{0}$. The Little-Parks oscillations of the resistance $\Delta R \propto  \overline{I_{p}^{2}} \propto  \overline{(n - \Phi/\Phi_{0})^{2}}$ have minimum at $\Phi = n \Phi_{0}$ and maximum at $\Phi = (n + 0.5)\Phi_{0}$ [9].

\section{Shift of the critical current oscillations measured on asymmetric rings}
The magnetic dependencies of the critical current $I_{c} = I_{c0} - 2|I_{p}|$ measured on symmetric ring [10] agree also with the persistent current oscillations (1) predicted with the universally recognized quantum formalism. The maximums of $I_{c}(\Phi/\Phi_{0})$ are observed at $\Phi = n \Phi_{0}$ and minimums at $\Phi = (n + 0.5)\Phi_{0}$ [10]. But we have discovered that the $I_{c+}(\Phi/\Phi_{0})$ and $I_{c-}(\Phi/\Phi_{0})$ oscillations of the critical current measured on asymmetric rings with different widths of the half-rings are shifted to opposite sides for measurement in the opposite direction [10]. This shift results to the anisotropy of the critical current $I_{c,an}(\Phi/\Phi_{0}) = I_{c+}(\Phi/\Phi_{0}) - I_{c-}(\Phi/\Phi_{0}) = I_{c}(\Phi/\Phi_{0}+\Delta \phi/2) - I_{c}(\Phi/\Phi_{0} -\Delta \phi/2) \neq 0$ explaining the rectification effect [8] observed on asymmetric superconducting rings [7]. But the observation of the $I_{c+}(\Phi/\Phi_{0})$, $I_{c-}(\Phi/\Phi_{0})$ maximums at $\Phi = (n \pm  \Delta \phi/2) \Phi_{0}$ and their minimums at $\Phi = (n + 0.5 \pm  \Delta \phi/2)\Phi_{0}$ is in an irreconcilable contradiction with the prediction (1) of the universally recognized quantum formalism. Because of possible fundamental importance of this contradiction we have measured the shift observed on ring with different radius and on system with different number of rings. 

Our previous measurements [10] of single ring with radius $r \approx  2 \ \mu m$ and widths of the half-rings $w_{w} = 0.25 \ \mu m$, $w_{n} = 0.2 \ \mu m$; $w_{w} = 0.3 \ \mu m$, $w_{n} = 0.2 \ \mu m$; $w_{w} = 0.35 \ \mu m$, $w_{n} = 0.2 \ \mu m$; $w_{w} = 0.4 \ \mu m$, $w_{n} = 0.2 \ \mu m$ have shown that the value of the shift $ \Delta \phi \approx  0.5$ independently of the value of the ring anisotropy $w_{w}/w_{n}$ in the interval 1,25 - 2 and of the critical current value $I_{c}(T)$. More careful measurements have confirmed this result for the critical current $I_{c}(T) > 10 \ \mu A$. But at the higher temperature, where the $I_{c}(T)$ is lower, we have found a weak decrease of the shift value: $ \Delta \phi \approx  0.4$ at $I_{c}(T) \approx 8 \ \mu A$; $ \Delta \phi \approx 0.35$ at $I_{c}(T) \approx 4.3 \ \mu A$; $ \Delta \phi \approx 0.3$ at $I_{c}(T) \approx 2.5 \ \mu A$; $ \Delta \phi \approx 0.25$ at $I_{c}(T) \approx 0.6 \ \mu A$, Fig.1. A more visible decrease of the shift is observed at measurement of system of 110 rings with radius $r \approx  1 \ \mu m$: $ \Delta \phi \approx 0.3$ at $I_{c}(T) \approx 4.1 \ \mu A$; $ \Delta \phi \approx 0.16$ at $I_{c}(T) \approx 2.8 \ \mu A$; $ \Delta \phi \approx 0.1$ at $I_{c}(T) \approx 1.6 \ \mu A$; $ \Delta \phi \approx 0.08$ at $I_{c}(T) \approx 0.7 \ \mu A$, Fig.2. Only insignificant discrepancy is observed between the magnetic dependencies  $I_{c+}(\Phi/\Phi_{0})$, $I_{c-}(\Phi/\Phi_{0})$ of the critical current measured in opposite directions on system of 667 asymmetric ring with radius $r \approx  0.5 \ \mu m$, Fig.3   

\section*{Acknowledgement}
This work has been supported by the  Russian Foundation of Basic Research grant 08-02-99042-r-ofi, a grant "Possible applications of new mesoscopic quantum effects for making of element basis of quantum computer, nanoelectronics and micro-system technic" of the Fundamental Research Program of ITCS department of RAS and a grant of the Program "Quantum Nanostructures" of the Presidium of RAS.

\end{document}